\documentstyle[psfig]{aa}
\input epsf         
\begin{document}

\title{X-ray He-like ion diagnostics: New computations for photoionized plasmas. II: Influence of different parameters.}

\author{ Olivier Godet\inst{1}, Suzy Collin\inst{2}, Anne-Marie Dumont\inst{2}}

\offprints{Olivier Godet (Olivier.Godet@cesr.fr)}

\institute{$^1$CESR,  9 av. du Colonel Roche, 31028 Toulouse Cedex 4, France \\
$^2$LUTH, Observatoire de Paris, Section de Meudon, Place Jules Janssen, F-92195 Meudon Cedex}

\date{Accepted : 29/06/2004}

\titlerunning{X-ray He-like ion diagnostics : New Computations
for Photoionized Plasmas}
\authorrunning{O. Godet et al.}

\abstract{
In a previous article, with a new code designed for a photoionized
thick hot medium using a full transfer treatment of both lines and
continuum, we have studied the He-like emission of a medium
photoionized by an X-ray source. We considered the influence of
density, ionization parameter and column density. We stressed that
photoionized media are not uniform in temperature and we showed that
photoexcitation and large column densities should be considered to
avoid misinterpretation of the properties of the medium.  To complete
the possible situations observed in photoionized plasmas, we consider
here the influence of microturbulence due for instance to a velocity
gradient, of different element abundance, and of different ionizing
spectral distributions. All new situations are computed using the same
code, but with an improved version of our He-like atomic model. We give
values of the $G$ and $R$ ratios, the equivalent width, and the ionic
column densities with a greater accuracy than previously achieved. A
large microturbulence results in a decrease of the $G$ ratios for
large column densities, for which the He-triplet resonant lines are
optically thick.  On the contrary, the $R$ ratio is almost independent
of the dispersion velocity and the element abundances, as long as the
intercombination lines are optically thin. Through two 'universal
diagrams' not depending on any parameter, we also present a way to
determine the ionization parameter and the total column density of the
emitting regions from the temperature of the observed radiative
recombination continuum of a given ion and its column density.

}

\maketitle

\section{Introduction}

In the era of the new X-ray space observatories {\it XMM-Newton} and
{\it Chandra}, the complexity of the soft X-ray spectra of several
objects has been discovered. They exhibit tens of emission lines ,
in particular the He-like ion lines of the n=2 complex, which can be
used as potential diagnostics to probe the physical state and to
constraint the geometry of the emitting regions.

It is well known that in collisional media, the He-like ion line
ratios $R$ and $G$ are sensitive respectively to the density and the
temperature of the emitting medium (see e.g. \cite{gab69},
\cite{gab72}, \cite{gab73}). We recall that these ratios are defined as:
$$\rm R = \frac{\rm z}{\rm x+\rm y} \hspace{6mm} \rm G = \frac{\rm z+\rm x+\rm y}{\rm w}$$  
where z is the forbidden line ($1s^2\ ^1S\ -2s\ ^3S$), x and y are the
intercombination lines ($1s^2\ ^1S\ - 2p\ ^3P_{1,2}$) and w is the
resonant line ($1s^2\ ^1S\ - 2p\ ^1P$).

More recently, these calculations have begun to be extended to 
photoionized plasmas, for objects such as active galactic nuclei (AGN)
and X-ray binaries (see the references given in \cite{coupe},
called hereafter paper I).

Aware of the strong potential of the He-like diagnostics to probe
these objects, we have undertaken an extensive study of photoionized
media. In paper I, we concentrated on the influence of the density
($n_{\rm H}$), the ionization parameter ($\xi$) and the column
density ($CD$) on the structure of the emitting region and the
He-like spectrum. We computed a grid of models, encompassing a large
number of cases where X-ray lines are observed. We used our code,
'Titan', designed for photoionized thick hot media, which performs a
full transfer treatment of both lines and continuum (see below), and
takes into account all relevant physical processes in the X-ray range.
This previous study led to some important conclusions concerning
photoionized media:

\begin{itemize}
\item  
It is difficult to {\it define a single temperature for the line
emitting region, because a photoionized medium is not uniform in
temperature}.

\item 
The $G$ ratios are {\it almost never equal to the pure recombination
value}. For low values of $CD$, $G$ is smaller, owing to photoexcitation
of the resonance line, and for high values of $CD$, it is larger, owing
to photon destruction during the process of resonant scattering.

\item 
From the study of a single ion, a given value of the $G$ ratio and a
given equivalent width (EW) could exist for a small $CD$ and a small
$\xi$, or for a large $CD$ and a large $\xi$. Other features of the
spectrum must be used to resolve this uncertainty.
\end{itemize}

This second paper has two motivations. First, the results mentioned
above were obtained with a relatively simple He-like atomic model
allowing only a qualitative description of the He-like emitting region
within an accuracy of $50\%$ for the pure recombination case as well
as for the pure collisional case.  Aware of this problem, we
implemented additional levels to build a more elaborate He-like atomic
model for the C~V, N~VI, O\,VII and Fe~XXV ions and checked our
conclusions given in paper I.  Note that the accuracy obtained with
the previous atomic model is better than that which could be obtained
when using the escape probability approximation: the errors on the
line intensities and line ratios reach thus one order of magnitude for
thick media (\cite{dumont03}) and typically a factor two for
'moderate' $CD$ (\cite{collin}).

Secondly, we complete the study of photoionized media in considering
other possible situations: the implications of a microturbulent
velocity added to the thermal velocity within the plasma, and the
influence of the element abundances on the structure and emission
properties with the example of an over/underabundance of oxygen, an
important element in X-ray emitting and absorbing media (e.g. the
Seyfert 2 nuclei and the Warm Absorbers). Also, in paper I, we
initiated an investigation of the influence of the spectral
distribution using an ionizing power-law continuum with various
spectral indexes. Aware that a power law is a poor representation of
the continuum in AGN in an energy range extending from UV to the
X-rays, we decided to continue this work using incident continua
closer to the mean quasar spectrum. \cite{Laor} have obtained
from a compilation of radio quiet quasars a 'mean spectrum'. We use
this continuum from 13.6 eV to 100 keV, since we are not interested in
the continuum below 13.6 eV, which does not contribute to the
ionization of heavy elements. If it would extend in the optical and
infrared band, it could however contribute appreciably to the heating
at high density through free-free process, and to the cooling at high
ionization flux through Compton process.  It has a spectral index
$\alpha=1.77$ from 13.6 eV to 2 keV, and $\alpha=1$ above 2 keV. We
call it the 'AGN1' continuum, in reference to the 'AGN' continuum
defined in paper I as a simple power law with $\alpha=0.7$ from 13.6
eV to 100 keV. However, since it is well known that in many objects
the spectral index above 2 keV is close to 0.7, we also use an
'AGN2' continuum with $\alpha = 0.7$ above 2 keV.

In paper I we reviewed the different X-ray emitting photoionization
media observed presently with {\it XMM-Newton} and with {\it Chandra}:
Seyfert 1-2, X-ray binaries. Here we focus on the pure emission line
spectrum seen in reflection as in Seyfert 2, because such a spectrum
does not depend on the proportion of continuum hidden by the emission
region. To this aim we have given all the results (equivalent widths,
line ratios, spectra) as they would be observed in reflection.  A
consequence is that the EW are much larger than in paper I, where
they were given relative to the incident continuum.  Note that the EW
are here upper bounds, as some sources of the continuum transmitted or
scattered could be only partly seen, even in Seyfert 2 nuclei.

In the next section we review the computational method and give a
short description of the atomic model. Sections 3 and 4 are devoted to
a discussion of the influence of the various parameters on the
structure and the emission of the medium, and Section 5 summarizes the
results.

\begin{figure*}
\begin{center}
\begin{tabular}{cc}
\psfig{figure=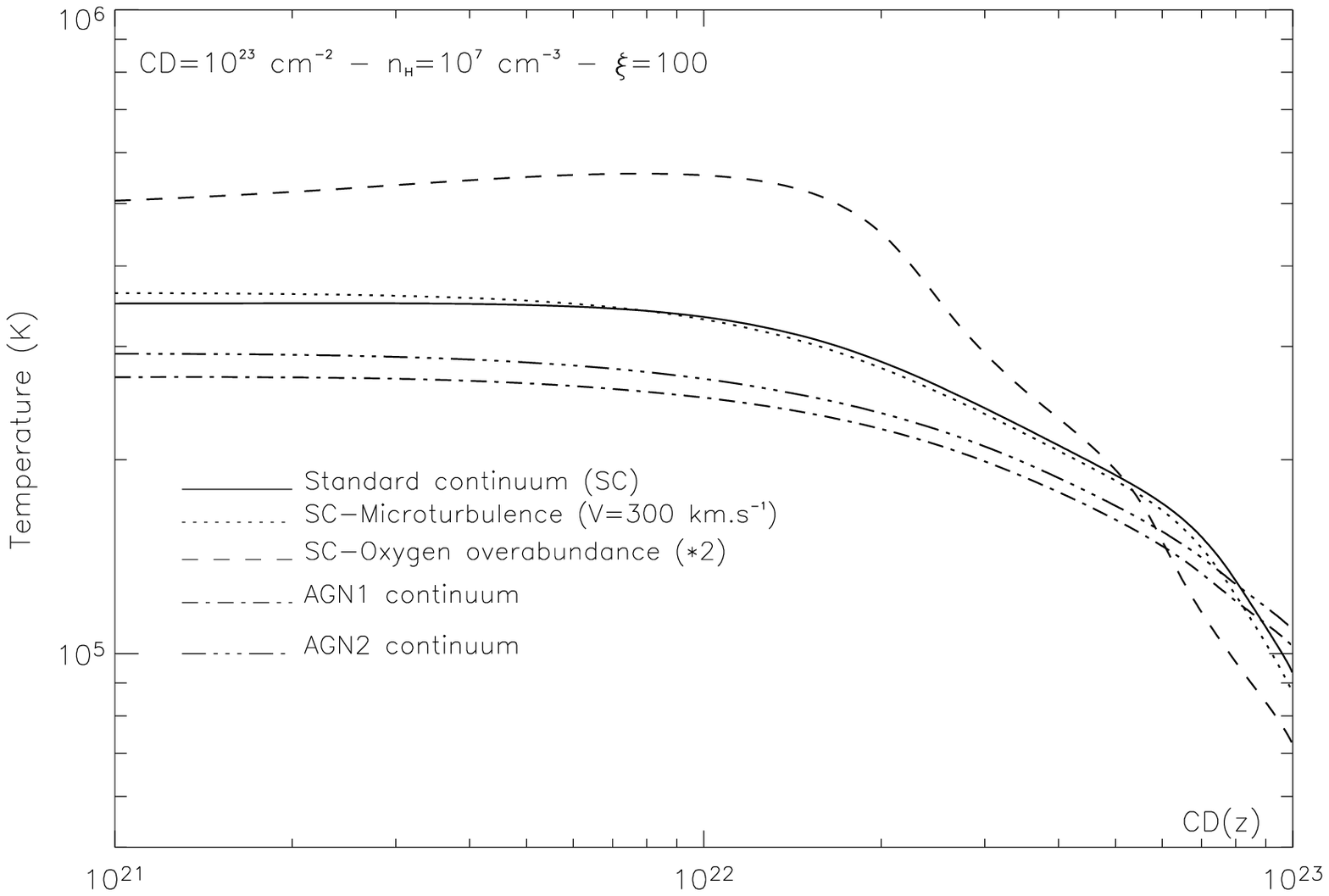,width=9cm,height=9cm} & \psfig{figure=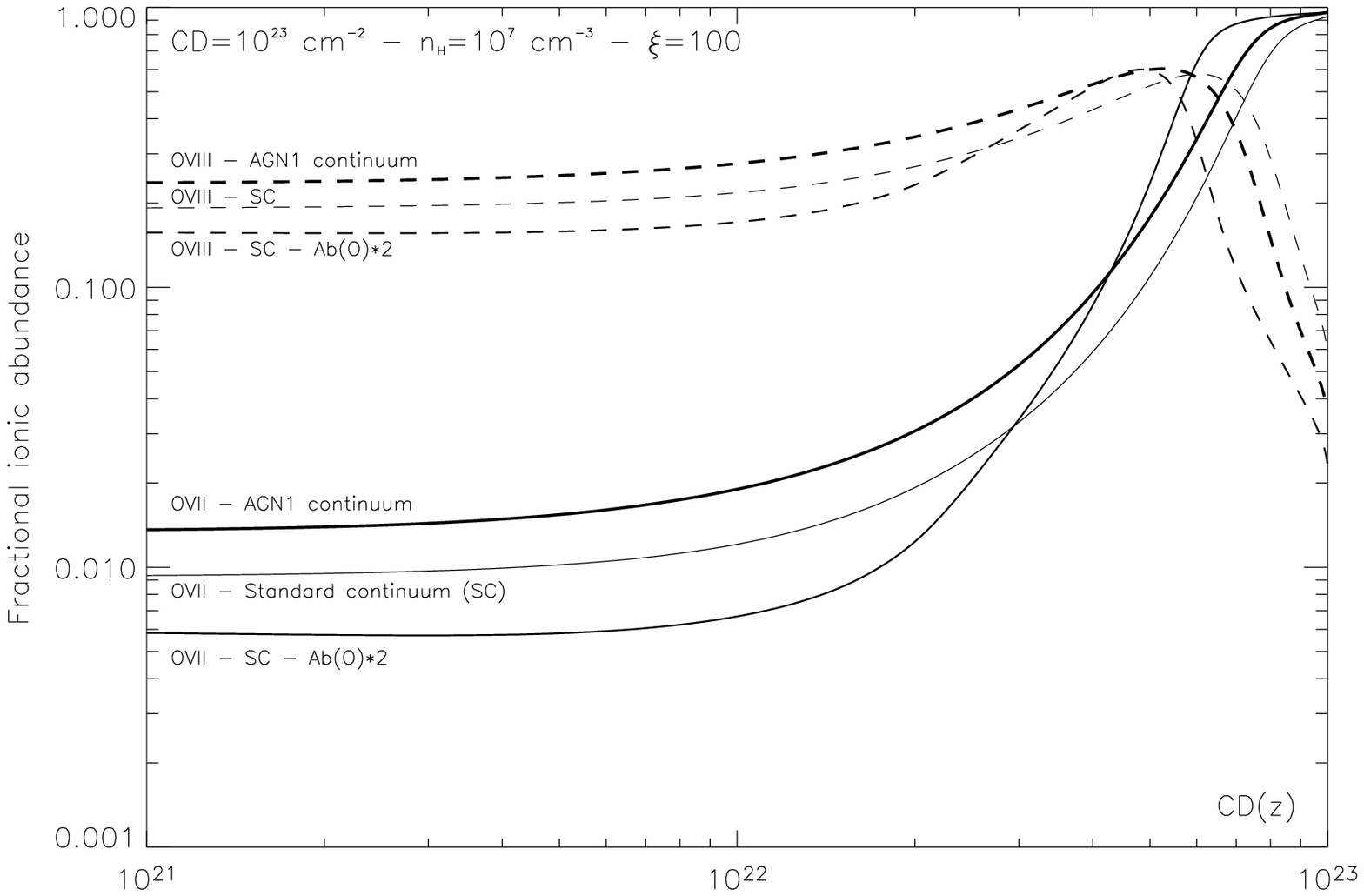,width=9cm,height=9cm} \\
\end{tabular}
\caption{Left panel: temperatures versus the column density $CD(z)$ for
$CD=10^{23}\,\rm cm^{-2}$, $n_{H}=10^{7}\,\rm cm^{-3}$ and
$\xi=100$. These temperatures are computed for several situations: the
new atomic data (solid line); microturbulence $V_{\rm
d}=300\,\mathrm{km.s}^{-1}$ (dotted line); twice the oxygen cosmic
abundance (dashed line); irradiation by the AGN1 and AGN2 continua
(dot-dashed line and 3 dot-dashed line respectively), the other ones
being computed with the standard continuum (SC). Right panel:
fractional ionic abundances of the O~VII and O~VIII ions versus
$CD(z)$ for the same model as mentioned above: twice the oxygen cosmic
abundance; irradiation by the AGN1 continuum. The labels on the curves
give the corresponding situations considered.}
\label{fig1x}
\end{center}
\end{figure*}

\section{The computational method}

\begin{figure*}
\begin{center}
\begin{tabular}{cc}
\psfig{figure=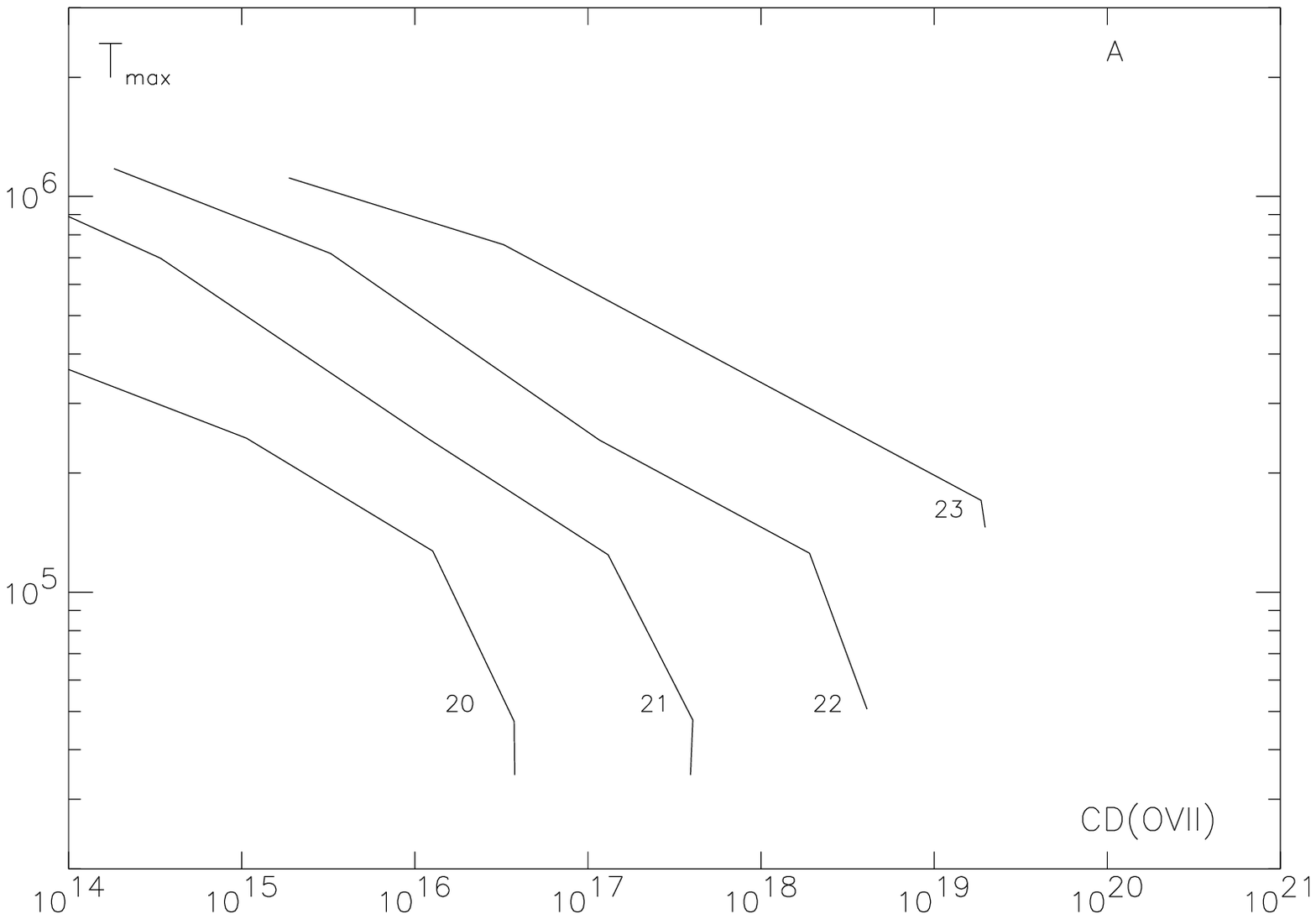,width=8cm,height=8cm} & \psfig{figure=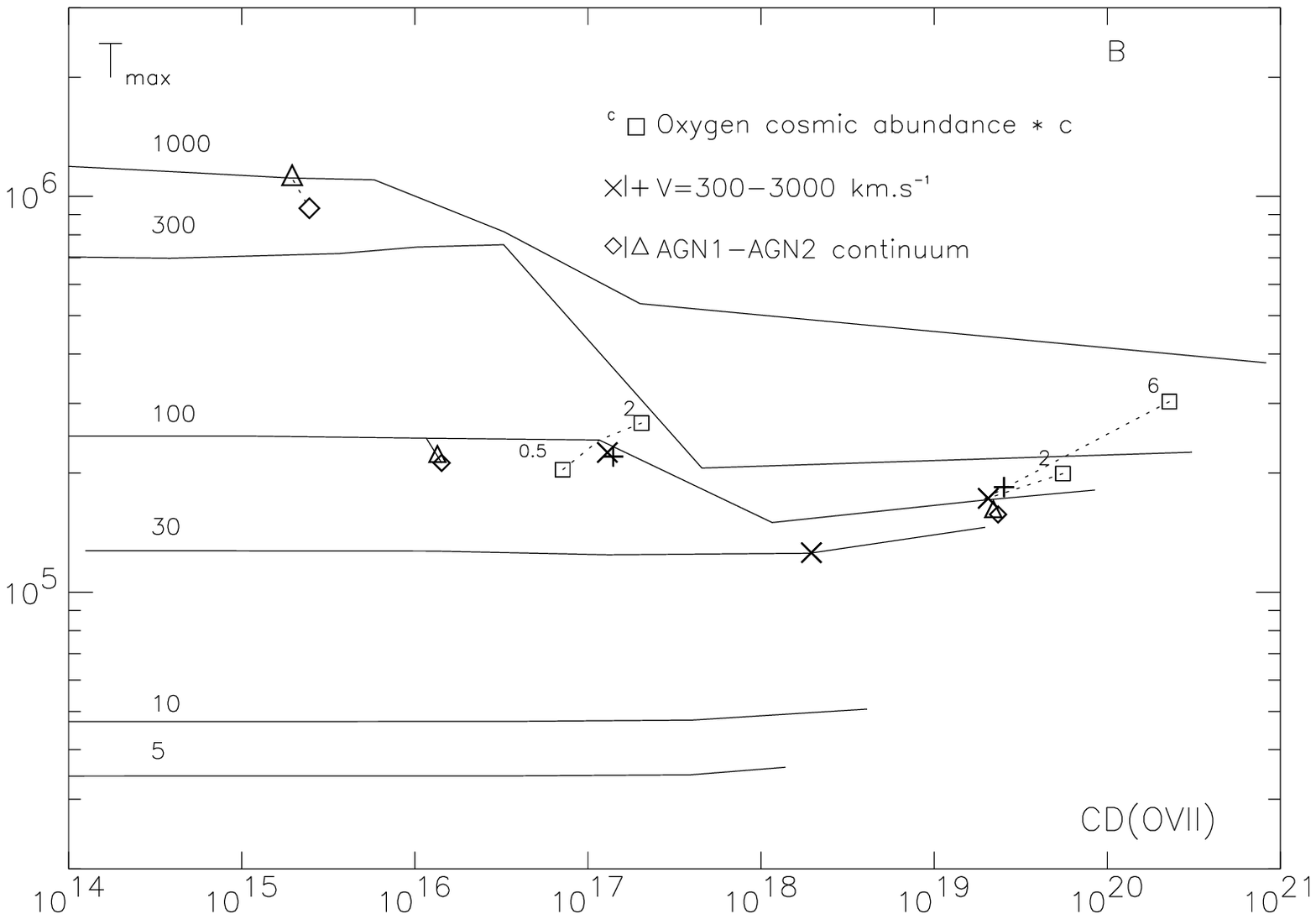,width=8cm,height=8cm} \\
\end{tabular}
\caption{ Maximum temperatures $T_{\rm max}$(K) of the regions where O~VIII 
dominate as a function of the O~VII column density (cm$^{-2}$)
according to: A- some given $CD$-values labelled on the curves in
logarithm; B- some given $\xi$-values labelled on the curves: the
symbol '+' and the crosses for $V_{\rm d}=3000-300\,\rm km.s^{-1}$
respectively; the squares for the models with a varying oxygen
abundance; the diamonds and the triangles for the AGN1 and AGN2
continua. The dotted lines in the right figure link models having the
same value of $\xi$ and $CD$. }
\label{fig2x}
\end{center}
\end{figure*} 

This work is performed with our stationary photoionization code
Titan used in paper I (for a detailed description of the
code, see \cite{dumont00}, \cite{dumont03}). Below, we give
a short description of the code and its originality with respect to
the other photoionization codes.

Titan uses the Accelerated Lambda Iteration (ALI) method to solve the
line and the continuum transfer. It is unprecedented from the point of
view of the line transfer for photoionization codes and allows us to
compute very precisely the line intensities, even for the most
optically thick lines. This cannot be performed with the other
photoionization codes like XSTAR (\cite{kallman}) and Cloudy
(\cite{ferland}), which use an integral method, the so-called
`escape probability' approximation, instead of solving the line
transfer and sometimes even the continuum transfer. With the new data
obtained with {\it XMM-Newton} and {\it Chandra}, it is crucial to use
a full transfer method as in Titan to properly compute the spectrum in
conditions appropriate for the X-ray emission regions of AGN.  Indeed,
it was shown in Dumont et al. (2003) and in Collin et al. (2004) that
the line intensities, mainly the resonant line, cannot be computed
properly with the `escape probability' approximation, owing to the
'photon destruction' process due to photoionization by the line photon
of ions having lower ionization potential for large $CD$, and to a bad
estimation of the photoexcitation rate of resonance lines by the
incident continuum for moderate $CD$.

A plane parallel slab of gas is irradiated by an incident spectrum,
whose frequency integrated flux is equal to $F$. We define the
`ionization parameter' $\xi = 4\pi\ F/ n_{\rm H}$
erg.cm.s$^{-1}$, where $n_{\rm H}$ is the hydrogen number density
in the plasma.  The gas composition includes the 10 most abundant
elements (H, He, C, N, O, Ne, Mg, Si, S, Fe) and all their ionic
species, i.e. 102 ions. A multi-level description is included up to
$n=5$ for H-like and Li-like ions, $n=4$ for He-like ions and $n=3$
for O IV and O V, where $n$ is the principal quantum
number. Interlocking between excited levels is included. Other ions
are treated more approximatively.

Titan includes all relevant physical processes from each level,
notably photoionization and photoexcitation, which is very important
to avoid misinterpretation of the He-like diagnostics, and all induced
processes.  The populations of each level are computed solving the set
of ionization equations coupled with the set of statistical equations
describing the excitation equilibrium, taking into account radiative
and collisional ionizations in all levels and recombination in all
levels, and radiative and collisional excitations and deexcitations
for all transitions. The energy balance is ensured locally with a
precision of 0.01\%, and globally with a precision of 1\%.

\subsection{The new He-like atom model}

In paper I, we used a model for He-like ions with only 11 levels and a
continuum. Then we implemented additional levels for the most
important He-like ions (C~V, N~VI, O\,VII and Fe~XXV). The sources of
the data are given in paper I, so we do not recall them here. The new
atomic model includes 15 levels and a continuum to take into account
all the terms for $n=2$ and $n=3$, rather than a single superlevel as
was previously done for $n=3$, and 2 superlevels gathering the singlet
and triplet levels for $n=4$. Recombination on the upper levels,
$n>4$, are taken into account and are distributed among the lower
levels according to their statistical weights. They represent at most
$50\%$ of the total recombination rates in all our
computations. Satellite lines are not included but they should be
negligible, like the dielectronic recombinations, in the range of
temperature where the important ions are present for the media
considered here (for instance $10^5$\,K for OVII, and $10^6$\,K for Fe
XXV).

\subsection{Models}

We ran models with a range of densities from 10$^{7}$ to 10$^{12}$
cm$^{-3}$, a range of column densities from 10$^{18}$ cm$^{-2}$ to
$3\times 10^{23}$ cm$^{-2}$ and a range of ionization parameters from
30 to 1000, with the new atomic data.  We assume cosmic abundances
with respect to hydrogen (\cite{allen}): He: 0.085; C: 3.3 10$^{-4}$; N:
9.1 10$^{-5}$; O: 6.6 10$^{-4}$; Ne: 8.3 10$^{-5}$; Mg: 2.5 10$^{-5}$;
Si: 3.3 10$^{-5}$; S: 1.6 10$^{-4}$; Fe: 3.2 10$^{-5}$.  We also
computed models with an oxygen abundance half, twice and six
times the cosmic abundance. The latter is an 'extreme' value,
certainly too high if we consider that the oxygen abundance is fixed
by stellar nucleosynthesis.  We also take into account the
possible existence of a microturbulent velocity (cf. Appendix for a
detailed explanation of what we call microturbulence) added to the
thermal velocity. We choose two values: $V_{\rm d}=300$ and $V_{\rm
d}=3000$ km.s$^{-1}$. The former is representative of the dispersion
velocity found in the X-ray spectrum of several Seyfert 1 and 2
galaxies (see e.g. \cite{O03}, and \cite{K02} for
\object{NGC 1068}). The latter is an 'extreme' value appropriate for
the broad absorption line region (BAL).  In paper I, the incident
spectrum was a power law, $F_\nu \propto \nu^{-1}$, extending from 0.1
eV to 100 keV. We will call it the 'standard continuum' (SC).  This
being a poor representation of the continuum in AGN, we also use the
AGN1 and AGN2 continua as mentioned earlier. We deal only with a
constant density inside the slab.

\section{Results}

\begin{figure*}
\begin{center}
\psfig{figure=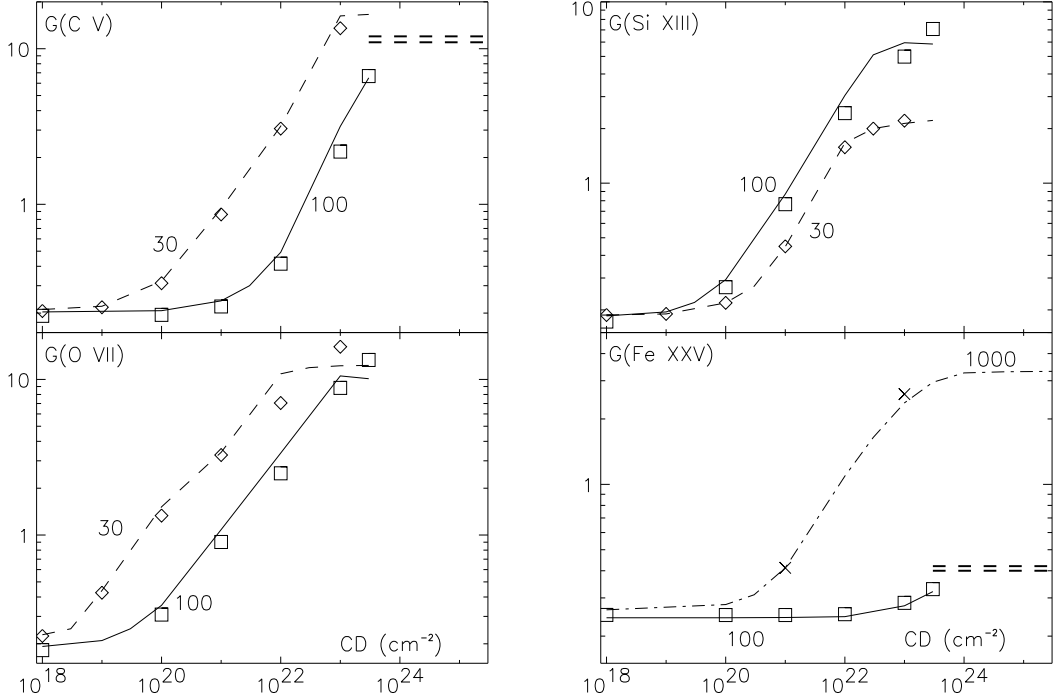,width=15cm,height=11.cm}
\caption{$G$ ratios in the reflected spectrum versus $CD$: Comparison between the new (the solid line: $\xi=100$; dashed line: $\xi = 30$) and the previous (squares: $\xi=100$; diamonds:  $\xi = 30$) atomic data for media with $n_{\rm H}=10^7$ cm$^{-2}$ irradiated by the standard continuum. For the Fe~XXV ions and the curve labelled at $\xi=1000$, the previous (dot-dashed line) and new (crosses) atomic data are shown. The double long dashed line for the C~V and Fe~XXV ions symbolizes the saturation of the $G$ ratio above a certain value of $CD$.}
\label{fig3x}
\end{center}
\end{figure*}

To show the different effects properly, we consider them separately. 

\subsection{New atomic data}  

First, it is interesting to consider whether the new atomic model for
C~V, N~VI, O\,VII, and Fe~XXV ions has a significant impact on the
thermal and ionic stratification of the medium. An important point to
bear in mind is that the location of the He-like emitting region
is closely linked to the stratification of the medium, as stressed in
paper I.  Indeed, we showed that a photoionized medium is divided into
several zones according to its thickness: a `hot skin' located below
the surface for which the temperature is roughly constant and the
elements are highly ionized, an `intermediate' zone where the
temperature decreases and the elements are less ionized, and a `cold'
zone where the elements are in low ionization states. In
Fig.~\ref{fig1x}, we show an example of the temperature and fractional
ionic abundances of the O~VII and O~VIII ions versus the column
density computed from the surface to the local point $CD(z)$ for a
medium with $CD = 10^{23}\, \rm cm^{-2}$, $n_{\rm H} = 10^{7}\, \rm
cm^{-3}$ and $\xi = 100$. It is appropriate for the Warm Absorber in
Seyfert 1 galaxies and the X-ray emission region in Seyfert 2
galaxies. For this moderate value of $CD$, only the `hot' and
`intermediate' zones could be seen. It is in this `intermediate' zone,
inhomogeneous in temperature and close to the back side of the slab, that
the O~VII spectrum is formed. A comparison with Fig.~2 of paper I
shows that the new atomic model of He-like ions almost does not modify
the fractional ionic abundance, and the new equilibrium in
temperature is globally close to the previous one with a relative
discrepancy of $10\%$.

In Fig.~\ref{fig2x}, several examples of the temperature $T_{max}$
of the O~VIII ions are given as a function of the column density of
the O~VII ions, according to some given $\xi$-values (panel B), and to
some given values of the total column density $CD$ (panel A).
$T_{max}$ is the maximum electron temperature in the region where the
O~VIII ions dominate.  We find that the curves with the new and old
atomic data are quite similar.  As stressed in paper I, this figure is
important, since it relates two quantities $T_{\rm max}$ (O~VIII) and
$CD$(O~VII) which can be deduced directly from the observed spectra,
$T_{\rm max}$ (O~VIII) corresponding to an upper bound of the
temperature derived from the O~VII radiative recombination continuum
(RRC).

The new He-like model allows us to get a better accuracy on the ratios
$R$ and $G$. Our previous model for the He-like ions gave $R$ correct
to only 20\%. We find a $G$-value in the pure recombination case for a
thin slab ($G=5.6$) for O\,VII ions similar to the value given in
Kinkhabwala et al. (2002), and a $R$-value for O~VII ions ($R = 3.9$
for $n_{\rm H} = 10^7\, \rm cm^{-3}$ and $\xi = 100$ \emph{i.e.} for a
temperature $T \sim 10^5\,\rm K$) similar to \cite{porquet00} in the
case of a very thin medium ($CD = 10^{18}\, \rm cm^{-2}$).

Fig.~\ref{fig3x} shows the $G$ ratios versus $CD$ for models with a
range of $\xi$ from 30 to 300 and $n_{\rm H} = 10^7\, \rm cm^{-3}$,
and a range of $CD$ from $10^{18}$ to $3\times 10^{23} \,\rm
cm^{-2}$. The new $G$-values fit well the previous $G$ ($5\%$) for
all ions except for OVII at large $CD$ at $\xi \le 30$. In this
figure, the long dashed line at $\xi=100$ symbolizes the saturation of
the $G$ ratio for the C~V and Fe~XXV ions. Note that $G$(O~VII)
saturates for smaller $CD$ at $\xi=30$.  We recall that this
saturation is due to the limit of the $CD_{ion}$ according to the
values of $\xi$, as shown in Fig.~4 in paper I.

\begin{figure}
\begin{center}
\psfig{figure=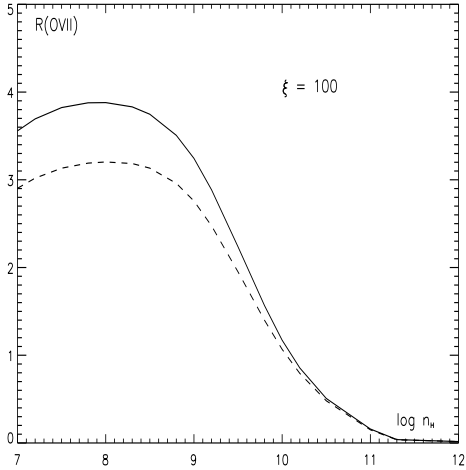,width=7.cm,height=7cm}
\vspace{-2mm}
\caption{$R$ ratio for the O~VII ion in the reflected spectrum versus the density $n_{\rm H}$ for media with $CD=10^{21}\,\rm cm^{-2}$, $\xi = 100$, photoionized by the standard continuum. The thick solid (dashed) lines correspond to the $R$-values computed with the new (old) atomic He-like model.}
\label{fig4x}
\end{center}
\end{figure}

Slightly more significant discrepancies between the previous and new
$R$-values can be seen in Fig.~\ref{fig4x}, showing $R$ for the O~VII
ion versus the density $n_{\rm H}$ for $CD = 10^{21}\,\rm cm^{-2}$ and
$\xi=100$.  At low densities, $R(\rm O~VII)$ computed with the new
He-like model is larger than the previous one by about 20\%, while at
high densities, the new $R$-values fit well the previous $R$-values.
This is because at high densities, the $n=2$ intercombination and
forbidden line transitions are basically driven by collisions. It
results in an increased forbidden line and a decreased
intercombination line.  Note that a similar behaviour is also seen for
the new $R$(C~V, N~VI)-values, while the new $R$(Fe~XXV)-values
present small differences with respect to the previous values ($\pm
2\%$).

In Fig.~\ref{fig5x}, we also give the EW computed as the sum of the
O~VII triplet lines with respect to the reflected continuum.

\subsection{Influence of microturbulence}

We take into account here the existence of microturbulence which,
unlike macroturbulence (like in the BLR of AGN), intervenes in the
line transfer, as explained earlier.

First, it is interesting to ask whether a microturbulent velocity has
a significant influence on the plasma temperature and on the line
intensities. Microturbulence decreases the optical thickness of the
lines at the line center, $\tau_0$, as it is inversely proportional to
$V_{\rm d}$. There is consequently an increase of emissivity of the
thick lines, as the rate of incident radiative excitation for these
lines increases when $\tau_0$ decreases, while the thin lines (for
instance the forbidden lines) are almost not affected, as stressed in
paper I.  Thus, the resonant lines w being optically thick for the
'moderate' and large $CD$, their intensities should increase when
$\tau_0$ decreases, while for the thin media where even the thickest
lines are optically thin, the line intensities do not vary
strongly. This is well illustrated in Fig.~\ref{fig6x} (panel B) which
shows the reflected spectra of models computed with the standard
continuum, $CD=10^{23}\,\rm cm^{-2}$, $\xi=100$, $n_{\rm H}=10^7\,\rm
cm^{-3}$, when comparing the panel B with $V_{\rm d} = 300\,\rm
km.s^{-1}$ to the panel A without microturbulence. All spectra are
convolved with a Gaussian profile of a FWHM equal to 1000 km.$\rm
s^{-1}$ (we recall that the lines can also be broadened by
macroturbulence).  As shown below, it will have also some consequences
on the $G$ ratios.

\begin{figure}
\begin{center}
\psfig{figure=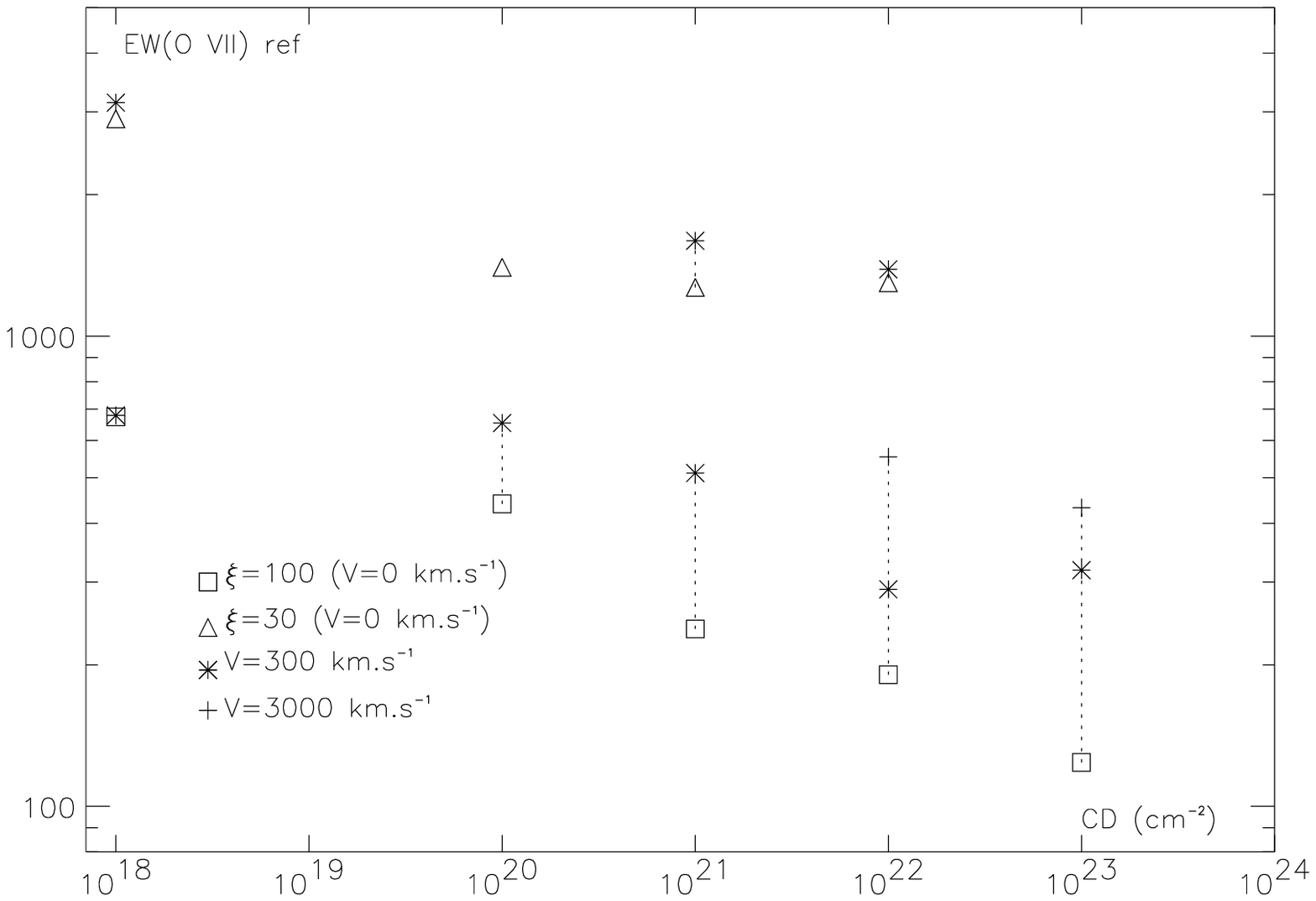,width=7.cm,height=7cm}
\vspace{-5mm}
\psfig{figure=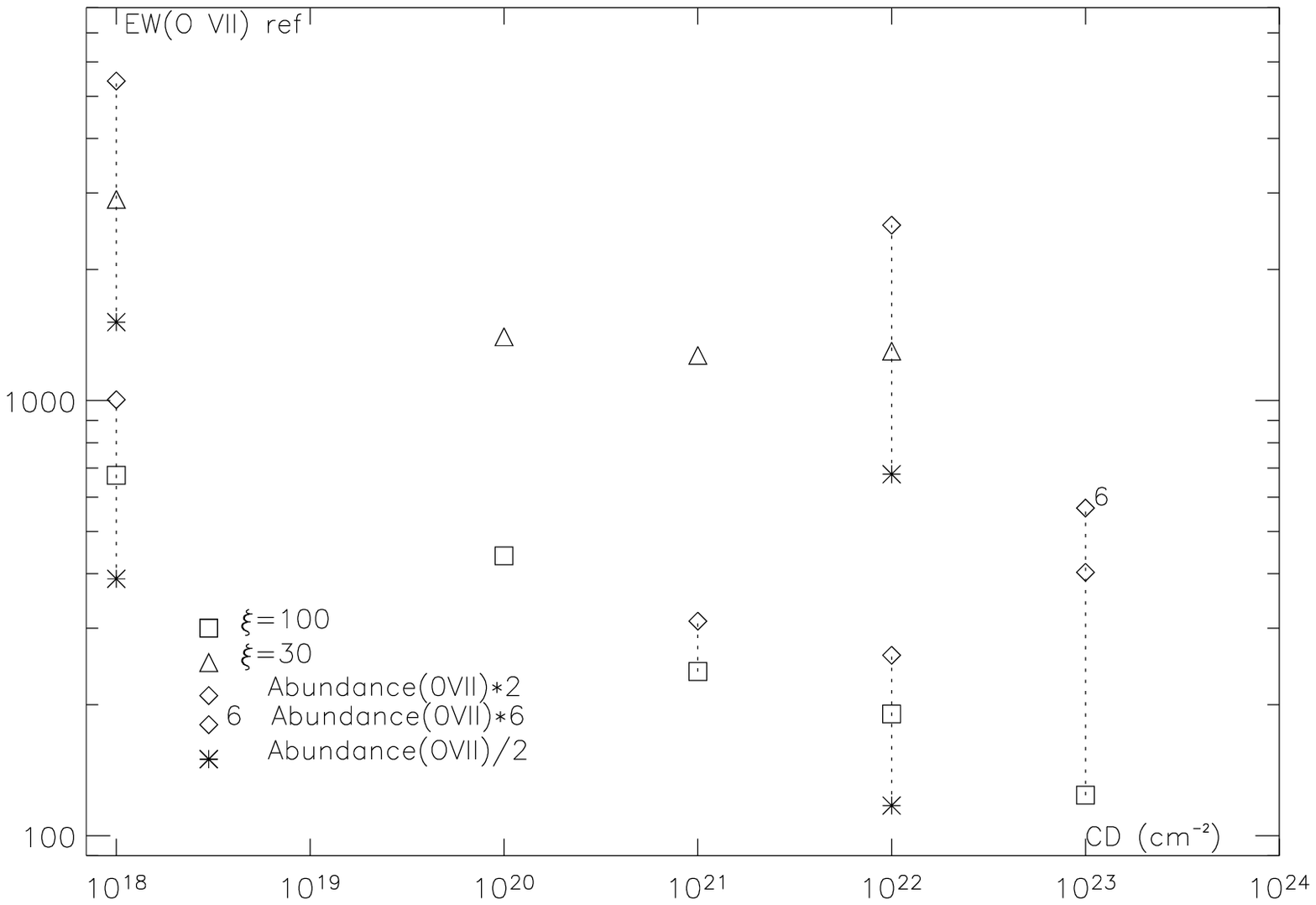,width=7.cm,height=7cm}
\caption{Reflected EWs of the sum of the O~VII triplet versus the column density of the slab. Top panel: models with $V_{\rm d}=300\,\rm km.s^{-1}$ (stars) and $V_{\rm d}=3000\,\rm km.s^{-1}$ (crosses). 
Bottom panel: models with a variation of the oxygen abundance:
abundance increased by a factor of 2 (diamonds). The diamond indexed
by '6' corresponds to an abundance increased by a factor of 6 for
$\xi=100$. The stars correspond to an abundance decreased by a factor
of 2.  For the two figures, the squares and the triangles corresponds
respectively to the reflected EWs for $\xi=100$ and 30.  The dotted
lines on the two figures link models having the same value of $\xi$.}
\label{fig5x}
\end{center}
\end{figure}    
\begin{figure}
\begin{center}
\psfig{figure=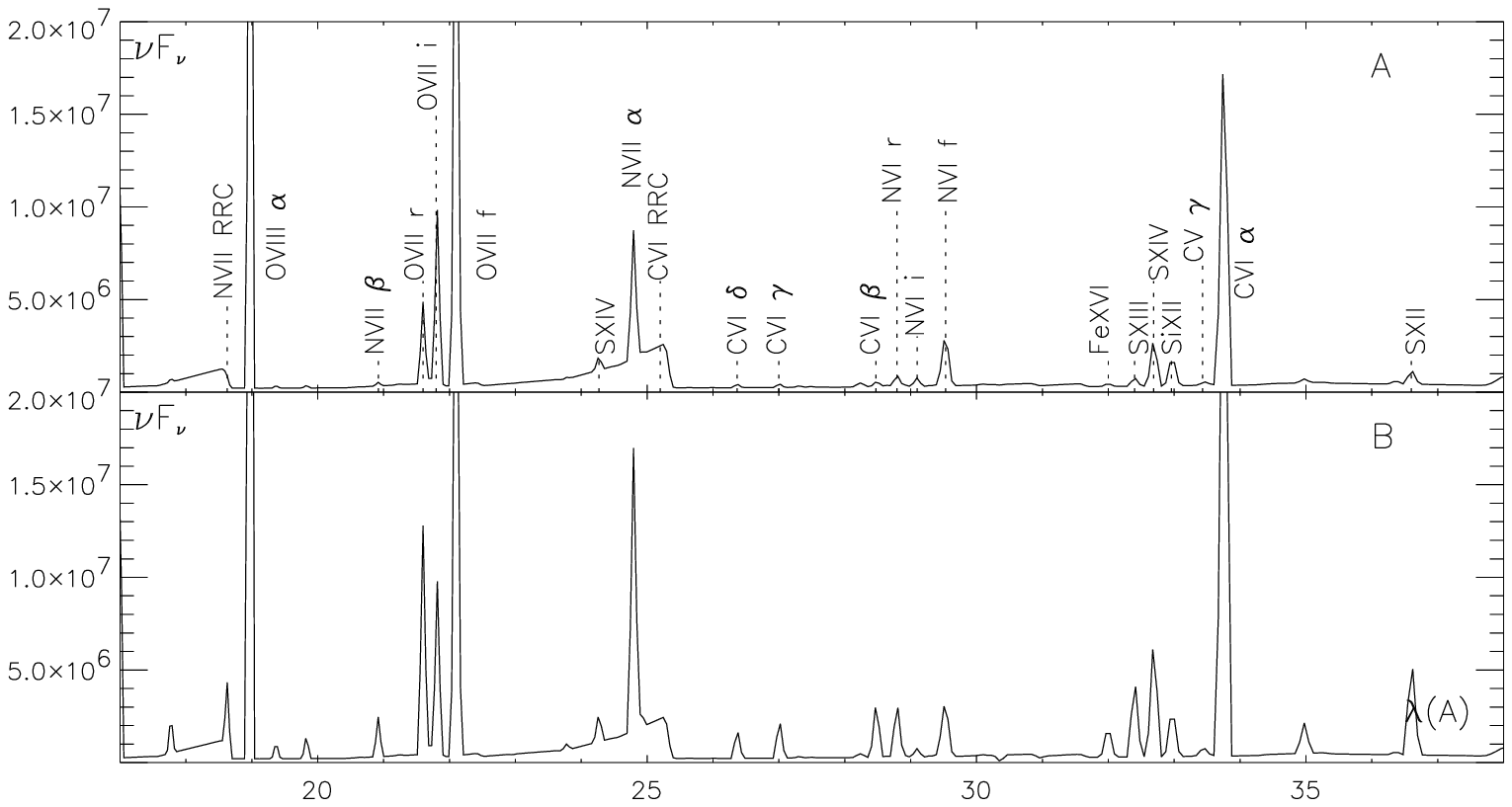,width=9.5cm,height=9.5cm}
\vspace{-3.2cm}
\psfig{figure=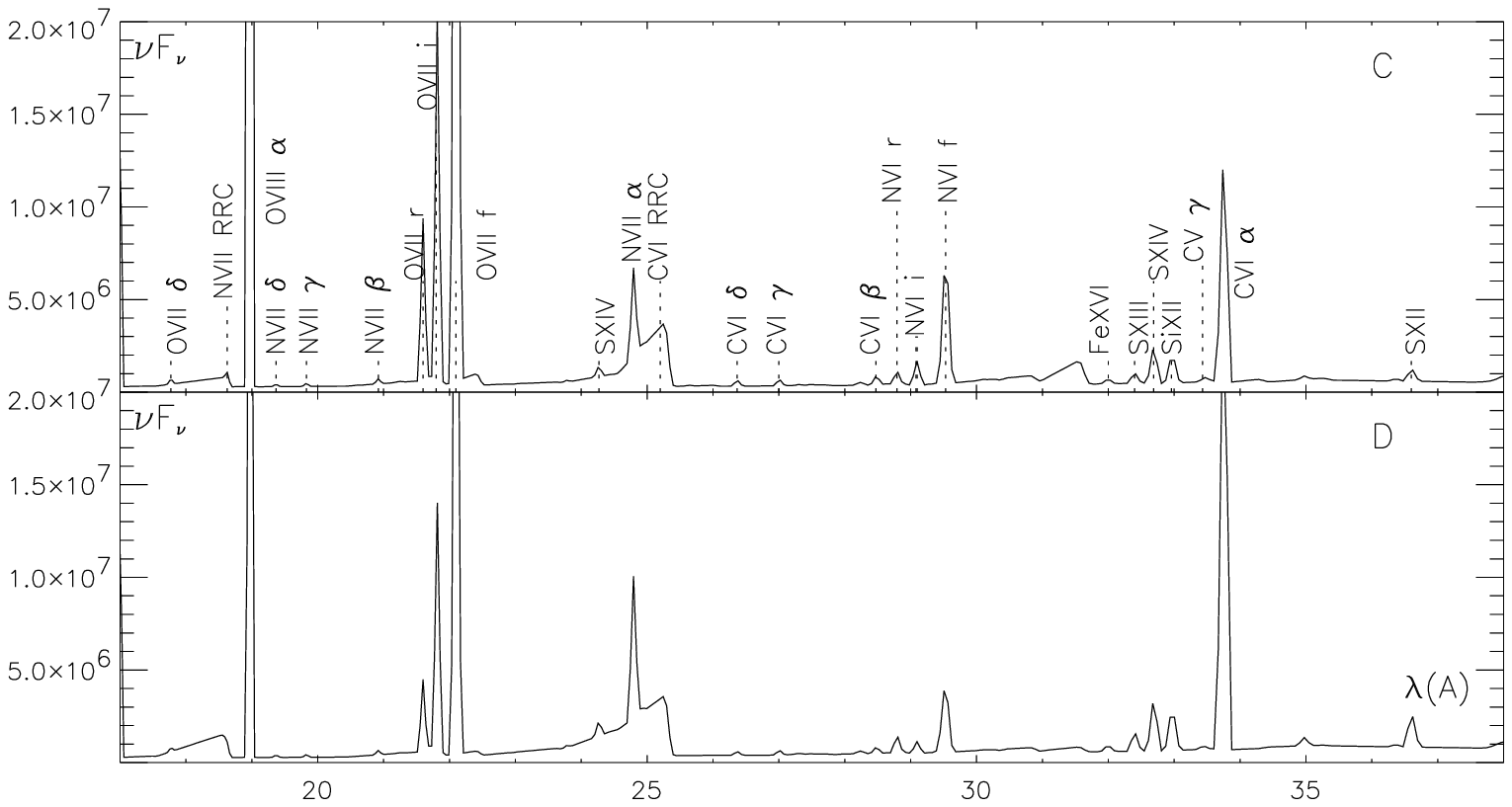,width=9.5cm,height=9.5cm}
\vspace{-3.5cm}
\caption{
Examples of reflected spectra given in a part of the RGS energy range
for $CD$=10$^{23}\,\rm cm^{-2}$, $n_{\rm H}=10^7\,\rm cm^{-3}$, $\xi =
100$ photoionized by the SC continuum. A- no microturbulent velocity
$V_{\rm d}$ and cosmic abundances; B) $V_{\rm d}=300\, \rm km.s^{-1}$
and cosmic abundances; C- no microturbulent velocity and an oxygen
abundance six times the cosmic abundance; D- no microturbulent
velocity and cosmic abundances, but irradiated with the AGN1
continuum.  The label 'i' corresponds to the sum of the two
intercombination lines x and y. All the spectra are convolved with a
Gaussian profile of a FWHM equal to 1000 km.s$^{-1}$. }
\label{fig6x}
\end{center}
\end{figure}

Microturbulence has a small influence on the temperature. For
a 'moderate' medium with $CD=10^{23}\,\mathrm{cm}^{-2}$, the
temperature computed for $V_{\rm d}= 300\, \rm km.s^{-1}$ is very
similar to that computed without microturbulence (cf. Fig. 1). For
$V_{\rm d}=3000\, \rm km.s^{-1}$, it increases by less than 15\% at the
illuminated surface and decreases by less than 30\% at the back side
with respect to that computed without microturbulence.

The ionic fractional abundances do not vary strongly when compared to
the previous values, notably for O~VII and O~VIII ions. It implies
that the ionic column densities are close to those found without
microturbulence, and thus {\it the $T(O~VIII)/CD$(O~VII) diagrams in}
Fig.~\ref{fig2x} {\it are almost independent of the microturbulent
velocity}.

\begin{figure}
\begin{center}
\psfig{figure=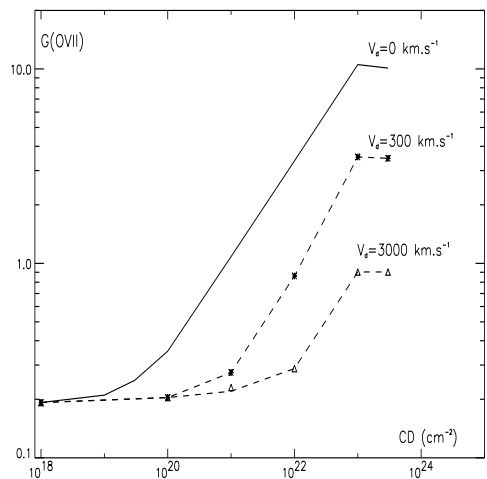,width=7.cm,height=7.cm}
\caption{$G$ ratio in the reflected spectrum for the O~VII triplet versus $CD$ for a medium with $n_{\rm H} = 10^7\, \rm cm^{-3}$, $\xi = 100$, photoionized by the standard continuum. The solid line with squares represents models without microturbulence; the dashed line with stars and the dashed line with triangles represent models with a microturbulent velocity of $300$ and $3000\, \rm km.s^{-1}$ respectively.}
\label{fig7x}
\end{center}
\end{figure}

Fig.~\ref{fig7x} shows the influence of the microturbulent velocity on
the $G$ ratios for the O~VII ions with $n_{\rm H}=10^7\,\rm cm^{-3}$
and $\xi=100$. $G$ decreases when $V_{\rm d}$ increases for a given
$CD$, except for $CD \le 10^{18}\,\rm cm^{-2}$ for the reasons
explained above. Note that $G$ saturates at the same column density
with and without microturbulence, but at smaller and smaller values,
as $V_{\rm d}$ increases. Therefore, for the same $G$-value,
microturbulence implies a larger CD than that derived from models
without microturbulence. Indeed, taking the example of \object{NGC
1068 given in paper I where $G(\rm O~VII)\sim 2.5$ (see \cite{K02}),
this value could be explained with $\xi\sim 100$ and $CD
\sim 10^{22}\,\rm cm^{-2}$, in the absence of microturbulence. For
$V_{\rm d}=300\,\rm km.s^{-1}$, we find that this same value can be
explained with $\xi \sim 100$ and a larger $CD\sim\,10^{23}\,\rm
cm^{-2}$.}

The results are however not trivial, as they depend critically on
the transfer treatment at moderate and low $CD$.  {\it Since, for
small $\xi$ ($\xi \leq 10$) and small $CD$} ($CD=10^{18-19}\,\rm
cm^{-2}$), {\it the optical thickness of the C~V and O~VII resonant
lines is still greater than unity, microturbulence has an
influence}. For instance, for a medium with $CD = 10^{18}\,\rm
cm^{-2}$, $n_{\rm H} = 10^7\,\rm cm^{-2}$, $\xi=5$, and $V_{\rm
d}=300\,\rm km.s^{-1}$, the $G$-values for these ions decrease by
$\sim 40\%$ with respect to those computed without $V_{\rm d}$.

The $R$(O~VII)-values with $V_{\rm d}=300\, \rm km.s^{-1}$ are similar
to those found without microturbulent velocity, because $R$ is, by
definition, the ratio of the forbidden lines over the intercombination
lines, which are optically thin. However, for the 'moderate' and thick
media it can happen that for some $\xi$-values the intercombination
line (x) of some ions is no longer optically thin, the y and z lines
being still optically thin, and $R$ then depends on the microturbulent
velocity.

Fig.~\ref{fig5x} (top panel) also shows the influence of
microturbulence, displaying the reflected EW of the sum of the O~VII
triplet.

\subsection{Oxygen overabundance and underabundance}

We limit ourselves to varying the abundance of the oxygen atom only,
as it is an important element in the spectrum of Warm
Absorbers-Emitters of Seyfert 1 nuclei and of the X-ray emission
region of Seyfert 2 nuclei. We recall that we consider an oxygen
abundance half, twice and six times the cosmic abundance.

Unlike microturbulence, varying the oxygen abundance induces a larger
variation in the temperature and the fractional abundances, as
displayed in Fig.~\ref{fig1x} for $CD=10^{23}\,\rm cm^{-2}$, $n_{\rm
H}=10^{7}\,\rm cm^{-3}$, $\xi=100$, and an oxygen abundance twice the
cosmic abundance.  The temperature at the surface irradiated by the
ionizing continuum is larger by $\sim 45\%$ than that computed with
the oxygen cosmic abundance, while the temperature of the back side
decreases by $\sim 25\%$.  Our interpretation is that heating near the
surface should be due to the direct X-rays absorbed on oxygen as well
as to the increased back-scattered radiation due to a larger number of
O~VII photons emitted by the back side. Cooling is not much affected
because the emission in the surface layers is dominated by the lines
from heavier elements, whose abundances are not changed. On the
contrary, at the back side, the O~VII cooling dominates and is
increased. Thus, the temperature of the back side decreases strongly,
but at the same time, heating near the surface causes an increase in
temperature.  The modifications of the He-like abundances are larger
for the light elements (C, N and O) than for the heavier elements (Si,
S, and Fe), as the light He-like elements are present in the same
region as oxygen, while the others are present closer to the surface.
For an 'extreme' value of six times the oxygen cosmic abundance, the
differences are more significant,with an increase of the temperature
by a factor of 2.5 at the surface, and a decrease of the temperature by
$70\%$ in the back side of the slab with respect to the values
computed with the cosmic abundance. This translates into differences
in the line intensities, in particular for the light elements. This is
well seen in Fig.~\ref{fig6x} (panel C) which displays an example of
the reflected spectra for $CD=10^{23}\,\rm cm^{-2}$, $n_{\rm
H}=10^{7}\,\rm cm^{-3}$, $\xi=100$ and an 'extreme' oxygen abundance
six times the cosmic abundance. Note that it is mainly the spectral
features of the light species that are modified, the heavy
species being almost the same, as shown above. Fig.~\ref{fig5x}
(bottom panel) also shows the impact of the oxygen abundance on the
reflected EW of the sum of the O~VII triplet.

As seen above in Fig.~\ref{fig2x} (panel B) with microturbulence, the
points with an oxygen under-/over-abundance do not strongly differ
from those obtained with the cosmic abundance, except for the extreme
increase of oxygen overabundance, where the discrepancies are more
significant (see the full circle in the panel B in Fig.~\ref{fig2x}).
{\it So again this diagram appears to be model independent}.

The $R$(O~VII) values are similar to those computed with the cosmic
abundance: the forbidden and intercombination lines being optically
thin, their intensities vary in the same way with abundance and $R$
consequently does not depend on the abundance.  However, for
'moderate' and large $CD$ and for some $\xi$-values, the
intercombination line, x, is optically thick, and $R$ depends on the
abundance. Thus, for a model with $CD=10^{23}\,\rm cm^{-2}$, $n_{\rm
H}=10^{7}\,\rm cm^{-3}$, $\xi=100$ and an oxygen abundance six times
the cosmic abundance, the $R$-value increases within 30\% of that with
the cosmic abundance.

\begin{figure}
\begin{center}
\begin{tabular}{c}
\psfig{figure=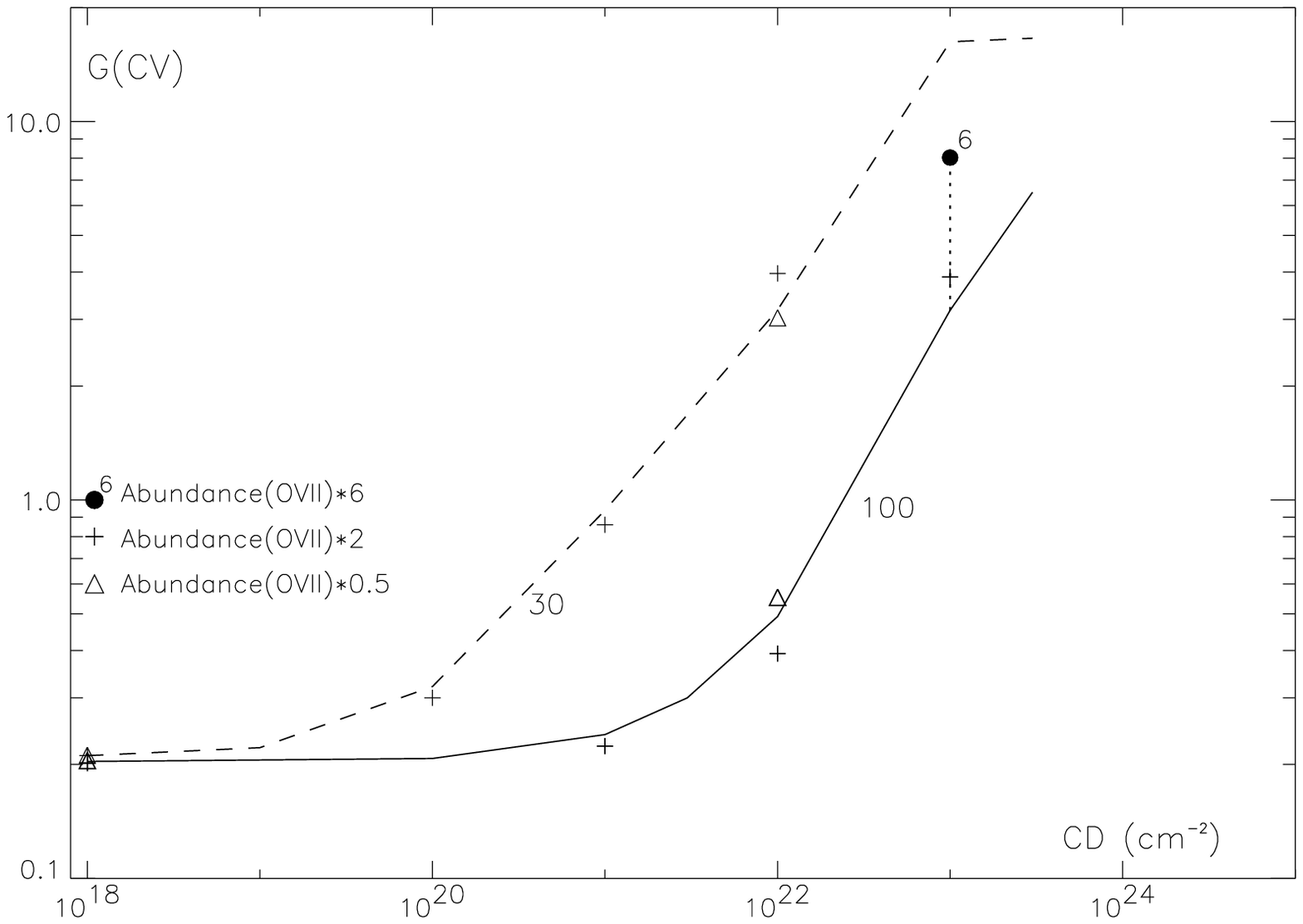,width=7.cm,height=7cm} \\
\psfig{figure=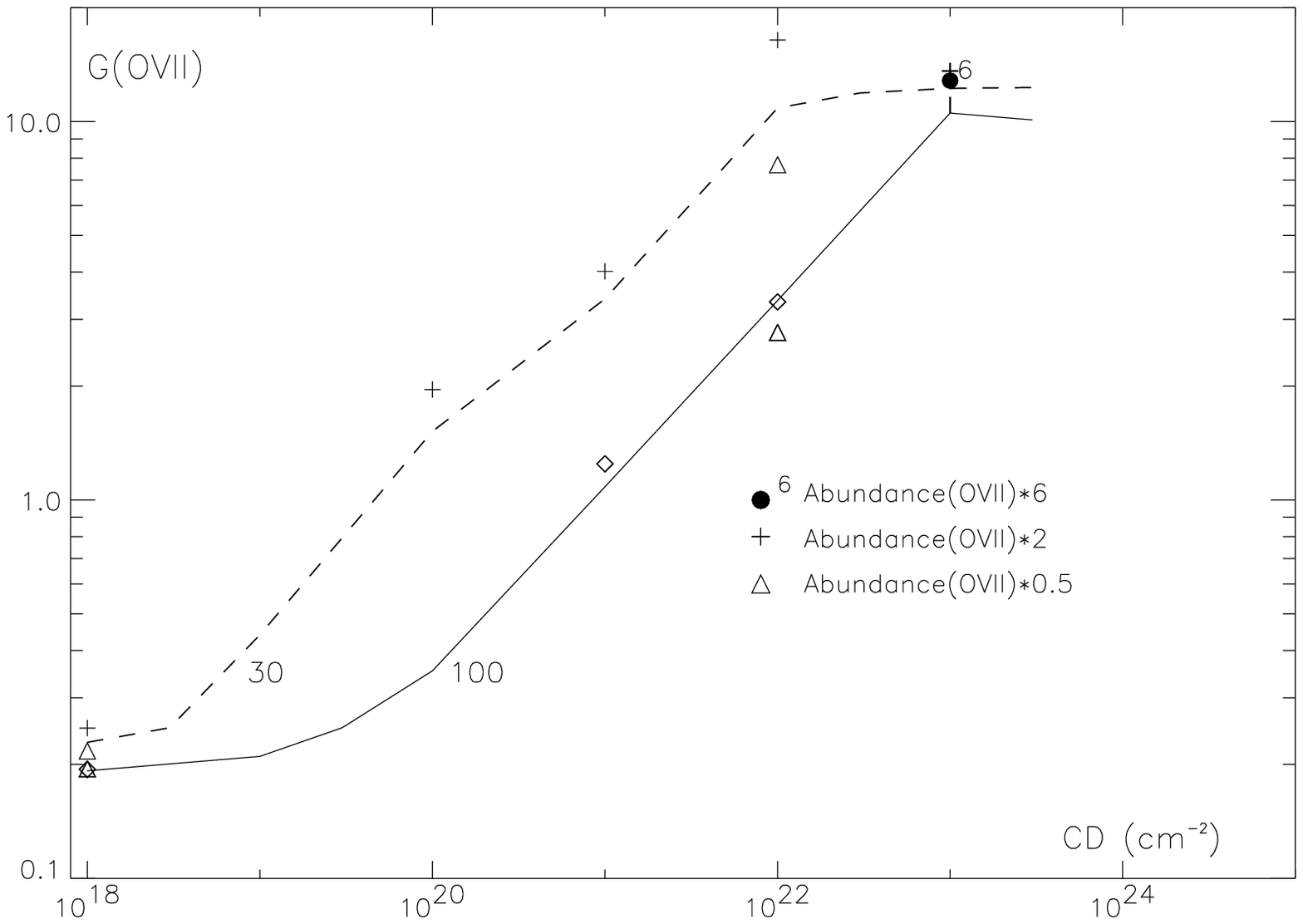,width=7.cm,height=7cm} \\
\psfig{figure=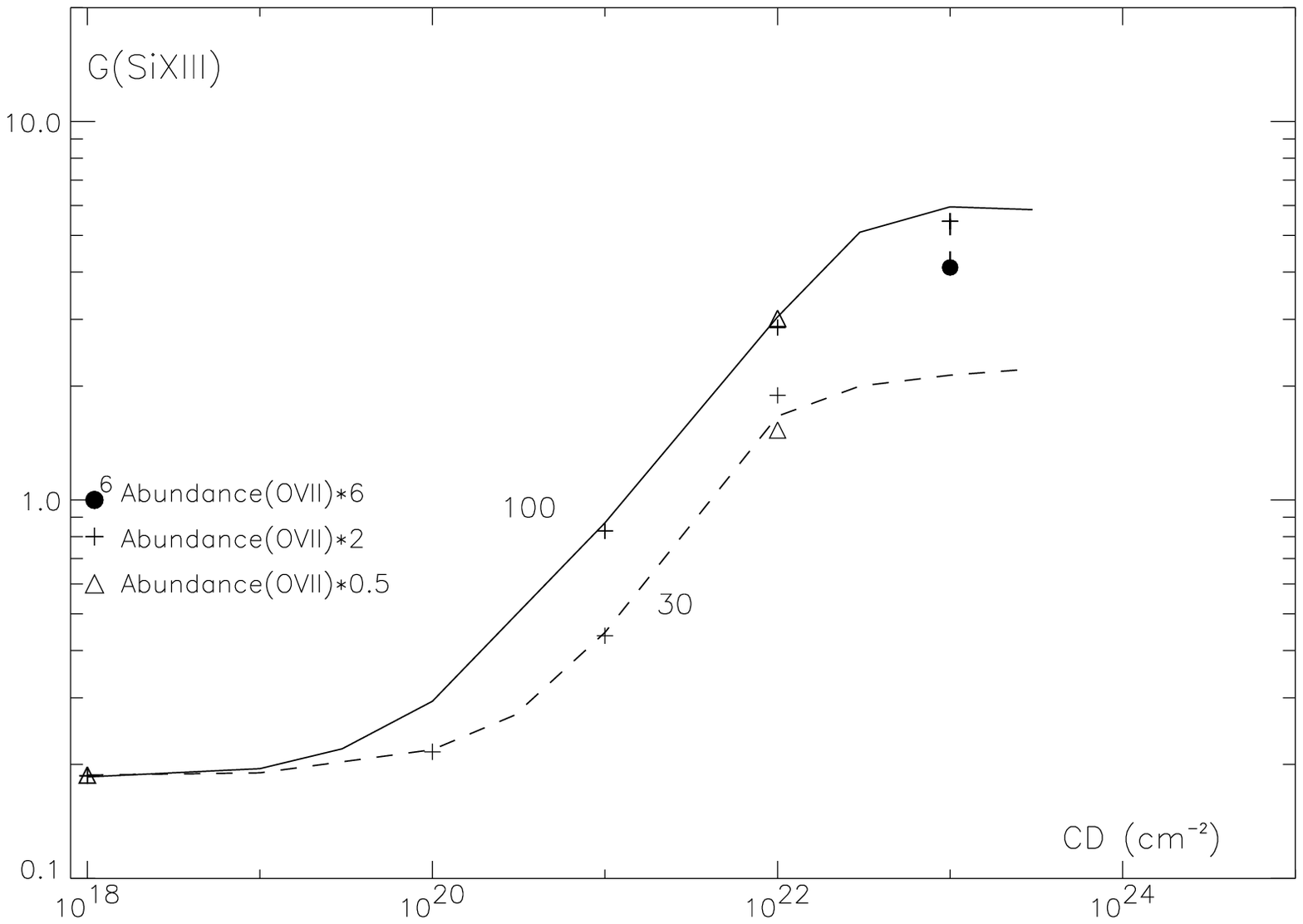,width=7.cm,height=7cm}  \\
\end{tabular}
\caption{$G$ ratio in the reflected spectrum for the C~V, O~VII and Si~XIII ions versus the column density for two values of the ionization parameter. The labels on the curves obtained with the oxygen cosmic abundance give the corresponding value of $\xi$. The triangles and the crosses correspond to an abundance 0.5 and 2 times the cosmic abundance for $\xi=30$. The stars and the diamonds correspond to the same for $\xi=100$. The full circle corresponds to an abundance 6 times the cosmic abundance for $\xi=100$.}
\label{fig8x}
\end{center}
\end{figure}

For the $G$ ratio, we see in Fig.~\ref{fig8x} that for very thin media
($CD=10^{18}$ cm$^{-2}$), the $G$(O\,VII) value is close to the value
computed with the cosmic abundance, because the resonant line is also
optically thin and so $G$ almost does not depend on the abundance. For
`moderate' and thick media, where the resonant and intercombination
(x) lines are optically thick and the forbidden and intercombination
(y) lines are still optically thin, $G$ depends on the abundance in a
complex way, as shown in Fig.~\ref{fig8x}.  However, this has only a
limited impact on the column density of the medium derived for all
elements. On the other hand, for an 'extreme' value (six times the
oxygen cosmic abundance), the $G$-values for the light ions (C~V and
N~VI) show some significant differences with respect to those found
with the cosmic abundance, as seen for the C~V ion in
Fig.~\ref{fig8x}, whereas for the heavy ions (e.g. Si~XIII) and O~VII
ions, $G$ presents some smaller differences (less than 30\%).

\begin{figure}
\begin{center}
\psfig{figure=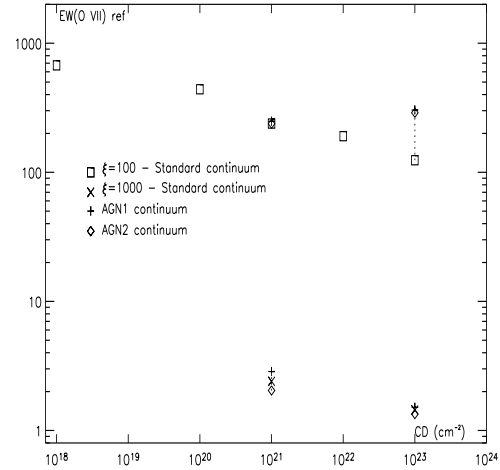,width=7.cm,height=7cm}
\caption{
Same caption as Fig.~\ref{fig5x} but for models irradiated by the AGN1
and AGN2 continua both normalized at 1 keV to the standard
continuum. The '+' and the diamonds correspond to the AGN1 and AGN2
continua respectively. The squares and the crosses correspond to the
values computed with the standard continuum, for $\xi=100$ and
$\xi=1000$ respectively.}
\label{fig9x}
\end{center}
\end{figure}

\subsection{The ionizing spectral distribution}
\label{sec34}

To compare the influence of the spectral distribution incident on the
medium, we normalized the different continua so that they have the same
flux at $1~\mathrm{keV}$.

The spectral distribution of the incident continuum has an influence
on the temperature and on the ionization state.  The left panel in
Fig.~\ref{fig1x} shows the temperature profile for a model with
$CD=10^{23} \rm cm^{-2}$, $n_{\rm H}=10^{7} \rm cm^{-3}$ and $\xi=100$
photoionized by the AGN1 and AGN2 continua. The temperature at the
surface is lower than that obtained with the standard continuum, while
the temperature near the back side is higher.  This is translated into
differences in the line intensities as we will see below. In the
right panel in Fig.~\ref{fig1x}, only the fractional abundances of the
O~VII and O~VIII ions obtained with the AGN1 continuum are given,
since the ionic stratification is globally close to that obtained with
the AGN2 models.

Note that in Fig~\ref{fig2x} the AGN1 and AGN2 points are
located relatively close to the points corresponding to the standard
continuum. Thus, {\it these diagrams are almost independent of the
ionizing spectrum}.

The panel D in Fig.~\ref{fig6x} shows the reflected spectrum for a
model with $ CD=10^{23}\,\rm cm^{-2}$, $n_{\rm H}=10^{7}\,\rm cm^{-3}$ and
$\xi=100$, irradiated by the AGN1 continuum.  It is similar to that
obtained with the AGN2 continuum, which is not displayed here.
Indeed, the temperature does not strongly differ between the AGN1 and
AGN2 models (cf. Fig.~\ref{fig1x}), so the line intensities are
similar. Both of these spectra present stronger radiative
recombination continua (RRC) for N~VII and C~VI, and stronger
C~VI$\,\alpha$, N~VI i and f, and O~VII i and f lines than the
reflected spectrum obtained with the standard continuum (see panel A
in Fig.~\ref{fig6x}).  We also give the reflected EW of the sum of the
O~VII triplet lines computed with the AGN1 and AGN2 continua in
Fig.~\ref{fig9x}. 

\begin{figure}
\begin{center}
\psfig{figure=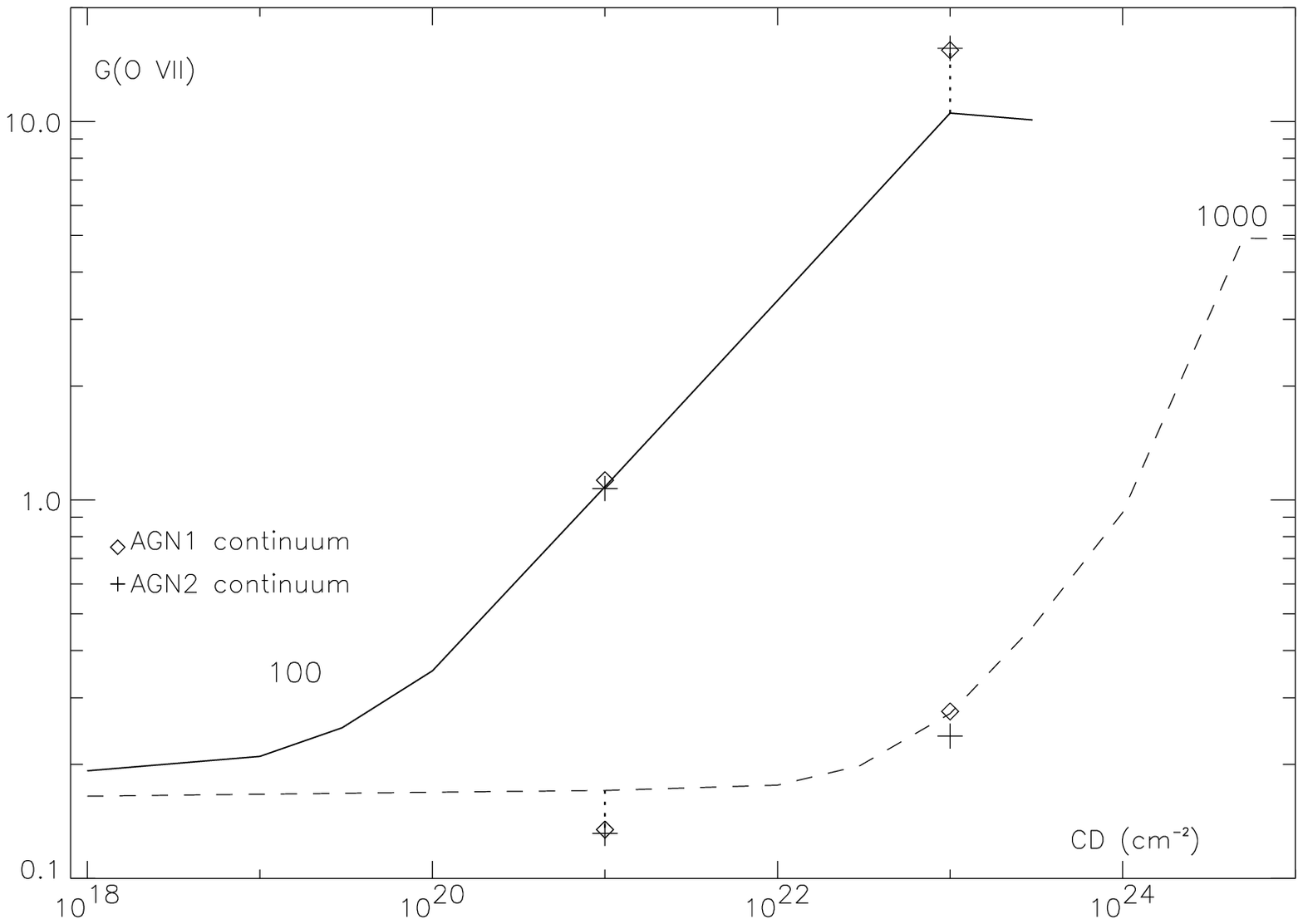,width=7.cm,height=7cm}
\vspace{-3mm}
\psfig{figure=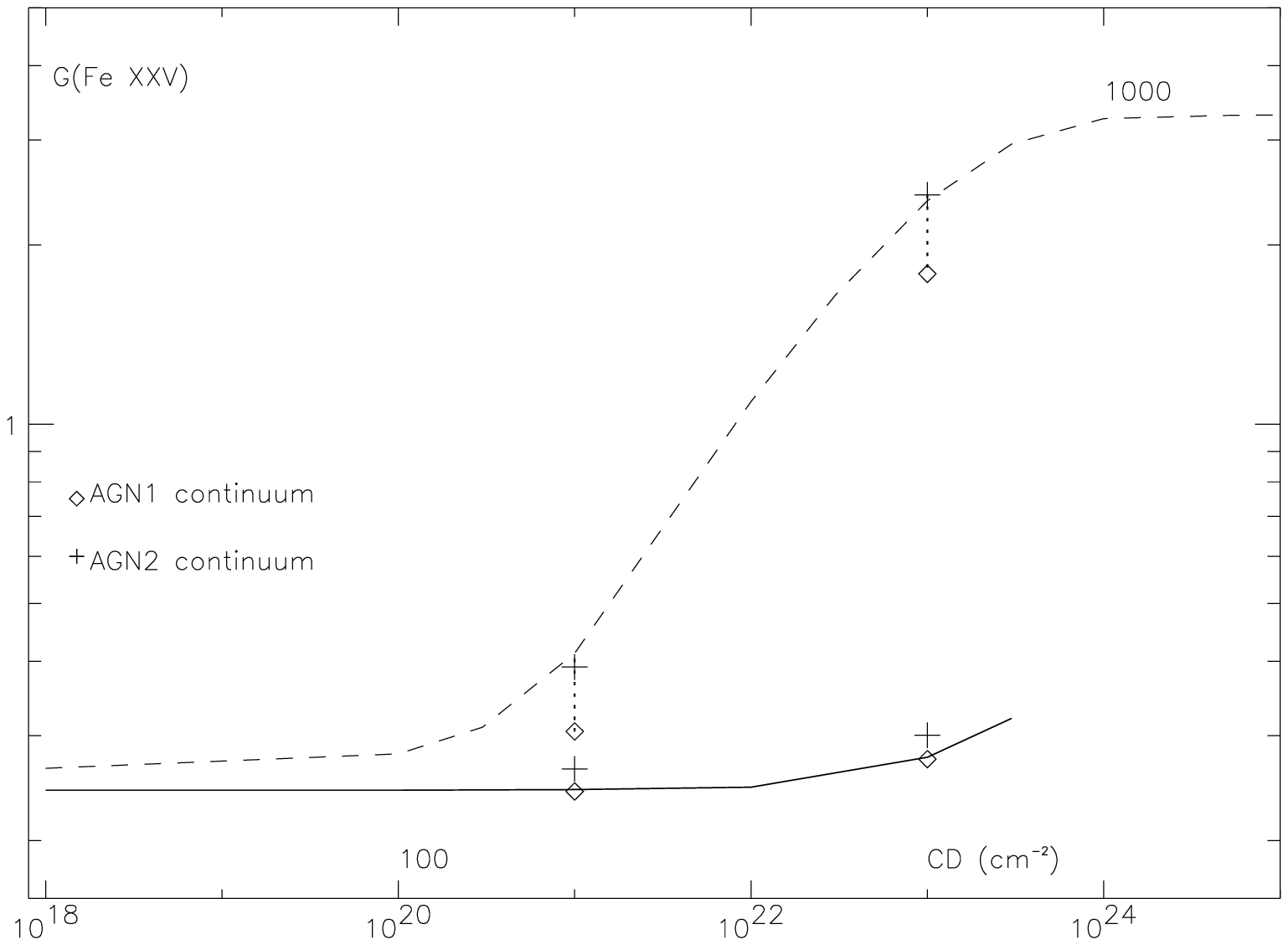,width=7.cm,height=7cm}
\caption{$G$ ratio in the reflected spectrum versus $CD$ for a medium with $n_{\rm H} = 10^7\, \rm cm^{-3}$, photoionized by the AGN1 and AGN2 continua both normalized at 1 keV to the standard continuum. The diamonds and the '+' symbols correspond to repectively the AGN1 and AGN2 continua. The dotted lines on the two figures link models having the same value of $\xi$. The curves are labelled with the value of $\xi$ for the standard continuum.} 
\label{fig10x}
\end{center}
\end{figure}

\begin{figure}
\begin{center}
\psfig{figure=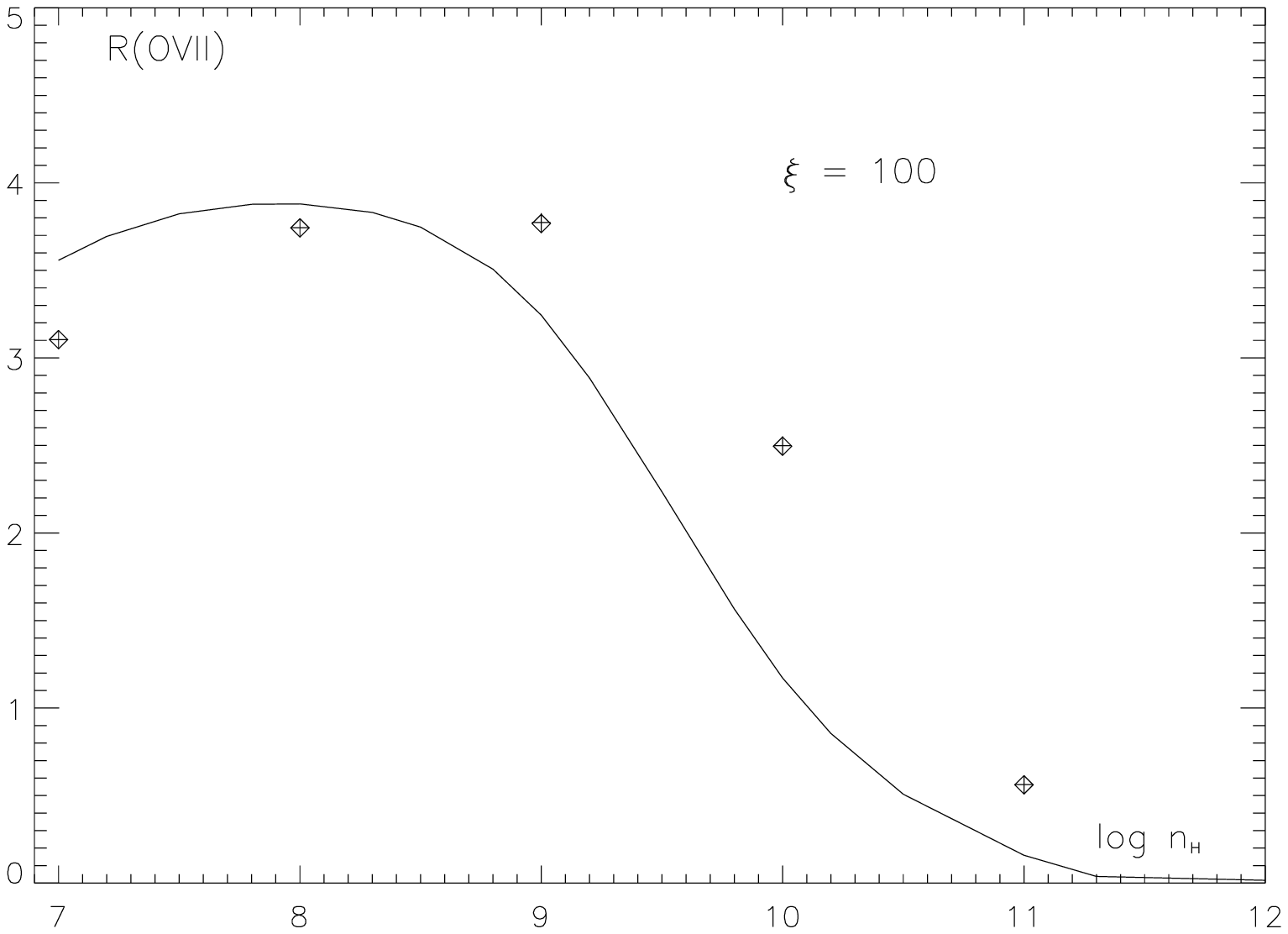,width=7.cm,height=7cm}
\caption{Same as Fig~\ref{fig4x}. The diamonds and the crosses correspond respectively to a model irradiated by the AGN1 and AGN2 continua (see the text for details).}
\label{fig11x}
\end{center}
\end{figure}

In Fig.~\ref{fig10x}, we give $G$ for the O~VII and Fe~XXV ions
computed with the AGN1 and AGN2 continua.  Note that for $\xi=1000$,
$G(\mathrm{Fe~XXV})$ presents some discrepancies between the values
computed with the AGN1 and AGN2 continua.  For $\xi=100$, the
$G$-values between the AGN1 and AGN2 continua are relatively similar
for all the ions.  The $R$ ratio, given in Fig.~\ref{fig11x} for the
O~VII ion, shows differences between the standard and AGN1-2
continua. These differences are significant in the region where $R$
decreases strongly with the density. In the region where $R$ is
approximatively constant, the $R$-values are close to those computed
with the standard continuum with differences less than $\sim 10\%$. We
observe the same behaviour for the heavy-Z Mg~XII, Si~XIV and S~XV
ions.  Note also that for all the ions, the $R$-values are very close
between the AGN1 and AGN2 continua for the $CD$ and $\xi$ values
considered here.

\section{Conclusion}

Using our new version of the photoionization code Titan and improved
atomic data for the He-like ions C~V-N~VI-O\,VII-Fe~XXV, we have
computed the line intensities emitted by an X-ray photoionized plasma.
We focused on the He-like emission for a range of ionization
parameters and column densities, considering several situations met in
photoionized objects. The results of this study are:

\begin{itemize}
\item 
The new atomic data of the He-like ions (C~V, N~VI, O\,VII and
Fe~XXV), taking into account all the separated $n=3$ levels, globally
confirms the results obtained with our previous He-like model. We find
that the new atomic data lead to $R$- and $G$-values similar to those
computed by other authors in the case of a pure recombination
medium. The $R$-values computed with the new atomic data differ by up
to 20\% from those previously computed at low densities.

\item
The different situations considered here show that {\it the diagrams
displaying the temperature of the O~VIII ion versus the ionic column
density of the O~VII ions are almost model independent}.Thus, if the
O~VII RRC temperature is derived from the observations, then it is
possible to determine a range of the $\xi$-values.

\item
We show that $R$ does not depend on microturbulence and on the oxygen
abundance, except for large column densities and low ionization
parameters.

\item 
The $G$ ratios of the He-triplets are very sensitive to the existence
of a microturbulent velocity, owing to the decrease of the optical
thickness of the resonant lines. This situation can be met in thin
media with low $\xi$, and 'moderate' and thick media with $\xi
\le 1000$. In these cases, for a given ionization parameter, $G$
decreases when the microturbulent velocity increases. In the other
cases, $G$ is independent of the microturbulent velocity, since all
He-like lines are optically thin.  Thus the column density deduced
from the observed G ratio is larger than that deduced without assuming
a microturbulent velocity. 
\end{itemize} 

Thus, our study has shown that in modelling X-ray spectra, it is
necessary to take into account the various situations met in the X-ray
emission regions of the different photoionized objects, such as
Seyfert 1 and Seyfert 2 nuclei, notably {\it the existence of a
microturbulent velocity is essential to avoid a misleading
interpretation from the He-like diagnostics}. The abundance study
shows once again the importance of taking into account the
interconnection between the different ions in the plasma although we
have considered here only the influence of the oxygen abundance.

The present results, computed with improved atomic data, are given
with an accuracy of the order of $\pm 10\%$. It is much better than
the accuracy obtained when using the escape probability approximation,
as all other photoionization codes do, whatever the sophistication of
the atomic model (see \cite{dumont03}, \cite{collin}).

Forthcoming papers will be devoted to building a grid of X-ray spectra
taking into account all these possibilities and applying it to the data
of the Seyfert 2 galaxies and X-ray binaries.

\vspace{1cm}
{\bf Appendix: On the relation between dispersion velocity and microturbulence}

The question of turbulent velocity is important, as it can strongly
modify the line ratios, but it is not simple. The effect of a
turbulent velocity on the line transfer can mimic that of a velocity
gradient or a dispersion velocity. However the simple existence of a
velocity gradient or a velocity dispersion does not necessarily mean
that it intervenes in the line transfer.  Let us recall why.

The ionization parameter can be written:
$$
\xi={R_{\rm Edd} L_{\rm Edd} f_{\rm vol} \over CD\ r \ R_G},
$$ where $R_{\rm Edd}$ is the bolometric to the Eddington luminosity
ratio $L/L_{\rm Edd}$, $f_{\rm vol} $ is the proportion of the line of
sight occupied by emitting material $(CD)/R$, i.e. the volumic filling
factor, and $r$ is the radius of this region expressed in the
gravitational radius $R_G$\,\footnote{We refer to $R_{G}$ instead of a
real dimension in parsec, as the locations and sizes of the different
regions around a massive black hole are all scaled by the mass of the
black hole. For instance, the radius of the BLR and of the Warm
Absorber in a Seyfert 1 nucleus is of the order of 10$^{4}R_G$, and
that of the mirror in a Seyfert 2 nucleus is of the order of
10$^{6}R_G$.}.

First one can show that the medium is not homogeneous in the Seyfert 
2 emission region and in the Warm Absorber of Seyfert 1.
 
 Assuming typical values for the Seyfert 2 emission region, one gets:
$$
\xi\simeq 66\ (R_{\rm Edd}/ 0.1)\ M_{8}^{-1}r_{6}^{-2} n_{5}^{-1}
$$ where $M_8$ is the black hole mass expressed in 10$^8$M$_{\odot}$
and $r_{6}$ the radius in $10^{6}R_{G}$, $n_{5}$ the density in
10$^{5}$ cm$^{-3}$. It shows that the spectrum would be dominated by
heavy He-like emission and not by OVII, unless the density is of the
order of or larger than 10$^{5}$ cm$^{-3}$.  For the Warm Absorber of
Seyfert 1, the density must be 10$^{9}$ cm$^{-3}$ or larger.

If $f_{\rm vol}$ would be equal to unity, one would get
$$ n\simeq 10^{3} CD_{22}M_{8}^{-1}r_{6}^{-1} $$
where $(CD)$ is expressed in 10$^{22}$ cm$^{-2}$. It is clearly too
small, implying that $f_{\rm vol}$ must be smaller than unity. A
similar conclusion can be deduced for the Warm Absorber. One gets for
the emission region of Seyfert 2: 
$$ f_{\rm vol} \simeq 10^{-3} (\xi/10)(R_{\rm Edd}/ 0.1)^{-1} CD_{22}r_6 $$

The emitting region is therefore made of small clouds, whose column
density $CD$ can be provided either by one cloud with a column density
equal to $CD$, or by a large number of smaller clouds.  Since the
dispersion velocity {\it inside} a cloud is of the order of the
thermal velocity, these clouds should have large {\it macroscopic}
velocities with respect to each others, in order to account for the
observed Doppler widths of the emission lines.

Let us consider the case of a unique cloud one a line of sight. The
surface coverage factor $f_{\rm surf}$ is thus: $$ f_{\rm surf}\sim
(N_c R_c^2)/R^2$$ where $N_c$ and $R_c$ are respectively the number
and the radius of the clouds. Since $R_c=(CD)/n$, one gets the maximum
number of clouds, which corresponds to a coverage factor equal to
unity: 
$$N_c^{\rm max}\sim 2\times 10^4 (M_8r_6n_5/CD_{22})^2$$ 

If the number of clouds is smaller than this value, $f_{\rm surf}$ is
smaller than unity. In both cases, a line emitted by one cloud would
not be absorbed by another one before escaping the medium. The
observed dispersion or gradient velocity would then be due only to the
respective motions of the clouds, and would not influence the transfer
like microturbulence does. This is the case in the BLR, as proved a long
time ago from the study of the line ratios.

Let us consider now the case where the individual size of the clouds
is smaller than $R_{\rm c}=(CD)/n$ and where there are several clouds
on a line of sight. Thus there must be $n_{ls}$ clouds, with
$R_c=n_{ls}R'_c$, where $R'_c$ is the new radius of the clouds. Thus a
line photon emitted by one cloud will cross other clouds before
escaping the medium, and it would not be reabsorbed since
the other clouds have large velocities corresponding to frequencies in
the wings of the line.  This is generally treated in the Sobolev
approximation. Here, we mimic this case by simply assuming a
microturbulent velocity equal to the macroscopic velocity of the
clouds with respect to each other, which amounts to assuming that the
optical thickness in all lines is so small that, once a photon has
been emitted in a cloud, it is never reabsorbed inside the same cloud
nor by the others. This means that $n_{\rm ls}$ should be larger than
the optical thickness at the center of the most optically thick lines
\emph{i.e.}  $n_{\rm ls}\ge 10^5$. The new surface coverage factor
$f'_{\rm surf}$ must be still of the order of unity in order to keep
the same line intensities. Since $f'_{\rm surf}$ is equal to $(N'_{\rm
c}/N_{\rm c}^{\rm max})/n_{\rm ls}^2$, where $N'_c$ is the new number
of clouds, we see that $N'_c$ must be larger than $10^{10} N_{\rm
c}^{\rm max}\sim 10^{14}$. Since we do not presently know the
structure of the emission medium, we have to consider both hypotheses.

\end{document}